\documentclass[english]{iopart}
				   
\begin{document}

\title[Beyond series expansions.]
{\Large
Beyond series expansions : mathematical structures for the 
susceptibility of the square lattice Ising model\footnote{
Dedicated to A.J. Guttmann on the occasion of his 60-th birthday.}.}
 
\author{ 
N. Zenine$^\S$, S. Boukraa$^\dag$, S. Hassani$^\S$ and
J.-M. Maillard$^\ddag$}
\address{\S  Centre de Recherche Nucl\'eaire d'Alger, \\
2 Bd. Frantz Fanon, BP 399, 16000 Alger, Algeria}
\address{\dag Universit\'e de Blida, Institut d'A{\'e}ronautique,
 Blida, Algeria}
\address{\ddag\ LPTMC, Universit\'e de Paris 6, Tour 24,
 4\`eme \'etage, case 121, \\
 4 Place Jussieu, 75252 Paris Cedex 05, France} 
\ead{maillard@lptmc.jussieu.fr, maillard@lptl.jussieu.fr, 
sboukraa@wissal.dz, njzenine@yahoo.com}

\begin{abstract}
We first study the properties of the Fuchsian ordinary differential equations for the
three and four-particle contributions $\, \chi^{(3)}$ and  $\, \chi^{(4)}$
of the square lattice Ising model susceptibility. An
 analysis of some mathematical  properties 
of these Fuchsian differential equations is sketched.
For instance, we study the factorization properties
of the corresponding linear differential
 operators, and consider the singularities of the 
three and four-particle contributions $\, \chi^{(3)}$ and 
$\, \chi^{(4)}$, versus the 
singularities of the associated Fuchsian ordinary differential
 equations, which 
actually exhibit new ``Landau-like'' singularities.
 We sketch the 
analysis of the corresponding
differential Galois groups. In particular we provide a simple, but efficient, 
method to calculate the so-called ``connection matrices''
 (between two neighboring singularities) and deduce the
 singular behaviors of $\, \chi^{(3)}$ and  $\, \chi^{(4)}$. 
We provide a set of comments and speculations on the Fuchsian
 ordinary differential equations 
associated with the $\, n$-particle contributions $\, \chi^{(n)}$ and address
the problem of the apparent discrepancy between such a holonomic approach and
some scaling results deduced from a Painlev\'e oriented approach.

\end{abstract} 
\vskip .5cm


\noindent {\bf AMS Classification scheme numbers}: 34M55, 
47E05, 81Qxx, 32G34, 34Lxx, 34Mxx, 14Kxx


\section{Introduction}
\label{intro}

It is a great pleasure to write this paper, in honour of the
 sixtieth birthday of Prof. A.J. Guttmann,
 on the mathematical
structures deduced from series expansion of the 
susceptibility of the Ising model.
One of us (JMM) has profited from many years of fruitful collaboration 
and correspondence with him and his group, and many times from his friendly
hospitality. 

\subsection{Notation and  
a few results on the Ising susceptibility.}
\label{recalls}



Since the (monumental) work of T.T. Wu, B.M. McCoy, C.A. Tracy and
E. Barouch \cite{wu-mc-tr-ba-76},
it has been known that the expansion in $n$-particle contributions to
the zero field susceptibility of the square lattice Ising model
at temperature $T$ can be written as an infinite sum :
\begin{eqnarray}
\label{coywu}
\chi(T)\,  =\, \,  \sum_{n=1}^{\infty} \chi^{(n)}(T) 
\end{eqnarray}
of $(n-1)$-dimensional 
integrals~\cite{nappi-78,pal-tra-81,yamada-84,yamada-85,nickel-99,nickel-00}.
This  sum is restricted to odd (respectively even) $n$ for the high
(respectively low) temperature case.
While the first contribution in the sum, $\chi^{(1)}$, 
is obtained directly without
integration, and the second one, $\chi^{(2)}$,
 is given in terms of elliptic integrals,
{\em no closed forms} for the higher order contributions were known,
 despite the well-defined forms of the integrands in these
 $(n-1)$-dimensional integrals.
 It is only recently that the differential equations for $\chi^{(3)}$
 and $\chi^{(4)}$ have been found
\cite{ze-bo-ha-ma-04,ze-bo-ha-ma-05,ze-bo-ha-ma-05b,zenine4,zenine5}.
In the following we will use, as in~\cite{ze-bo-ha-ma-04,ze-bo-ha-ma-05,ze-bo-ha-ma-05b,zenine4},
the following normalization :
\begin{eqnarray}
\label{normal}
\chi^{(n)} \, = \, \, {{(1-s^4)^{1/4} } \over {s}} \cdot \tilde{\chi}^{(n)}
\end{eqnarray}
in order to focus on the $\, \tilde{\chi}^{(n)}$'s which are only functions of the 
variable $\, w \, = \, s/(1+s^2)/2$. 

As far as  singular points are concerned (physical or 
non-physical singularities in the complex plane), 
and besides the known $\, s= \pm 1$ and $\, s\, = \pm i$ singularities, 
B. Nickel showed~\cite{nickel-99} 
that  $\, \chi^{(2\, n+1)}$ is singular\footnote[2]{The 
singularities being logarithmic branch points of order 
$\, \epsilon^{2\, n\, (n+1)-1} \cdot ln(\epsilon)$
with $\, \epsilon \, = \, 1-s/s_i$ where $\, s_i$ is one 
of the solutions of (\ref{sols}).
} for the following finite values of
$\, s=sh(2\,J/kT) \,$ lying on the $\, |s|\, =1$ {\em unit circle} ($m=k=0$ excluded):
 \begin{eqnarray}
\label{sols}
&&2 \cdot \Bigl(s \, + \, {{1} \over {s}}\Bigr) \, = \, \, \, \, 
u^k\, + \,{{1} \over {u^k}} \,
+ \, u^m\, + \,{{1} \over {u^m}} \,\nonumber \\
&&  \qquad u^{2\, n+1} \, = \, \, 1,  \qquad   \qquad  
-n \,  \,\le\,  \, m, \, \, k\,  \,\,\le \, \, n  
\end{eqnarray}
When $n$ increases, the singularities of the higher-particle components 
of $\chi(s)$ {\em accumulate on the unit circle} $\, |s|\, = \, 1$.
The existence of such a {\em natural boundary} for the total susceptibility $\,\chi(s)$, 
shows that  $\,\chi(s)$ is {\em  not D-finite} (not holonomic) 
as a function\footnote[3]{See in particular, the paper 
of I. Enting and A.J. Guttmann~\cite{gut-ent-96}. One has probably the same situation
for the {\em three-dimensional} Ising model~\cite{ha-ma-oi-ve-87}.} of $\, s$.

A significant amount of work had already been performed to
generate isotropic series coefficients
for $\chi^{(n)}$ (by B. Nickel \cite{nickel-99,nickel-00} up to order 116,
then to order 257 by 
A.J. Guttmann\footnote[9]{Private communication.} {\it et al.}).
More recently,  W. Orrick {\it et al.}~\cite{or-ni-gu-pe-01b,or-ni-gu-pe-01},
have generated coefficients\footnote[8]{The short-distance terms  were shown to
have the form $\, (T-T_c)^p \cdot (log|T-T_c|)^q\, $
with $\, p \ge q^2$. }
 of the susceptibility of the Ising model~\cite{or-ni-gu-pe-01b,or-ni-gu-pe-01}
 $\chi (s)$ up to order 323 (resp. 646) for high (resp. low) temperature 
series in $\, s$,
using  a quadratic double recursion  ({\em non-linear Painlev\'e difference equations})
for the correlation functions~\cite{or-ni-gu-pe-01b,or-ni-gu-pe-01,coy-wu-80,perk-80,jim-miw-80}.
As a consequence of this  remarkable non-linear Painlev\'e difference equation
and its  associated double recursion, the computer 
algorithm had a $\, O(N^6)$ {\em polynomial growth} 
for the calculation of the series expansion 
instead of the exponential growth
that one might expect at first sight.


However, in such a {\em non-linear}, non-holonomic,
{\em Painlev\'e-oriented} approach,
 one obtains results directly for the {\em total} susceptibility $\chi (s)$
 which {\em do not} satisfy any finite order linear differential equation. It thus prevents the easy 
disentanglement of the contributions of the various holonomic $\chi^{(n)}$'s. 

On the other hand, the individual $\chi^{(n)}$'s {\em do satisfy} finite order
linear differential equations, leaving some hope  
to understand the susceptibility
from a deeper knowledge of the mathematical structures of
the successive differential equations of the $n$-particle
 sequence of the $\chi^{(n)}$'s.


\section{The Fuchsian ODE's for $\chi^{(3)}$ and $\chi^{(4)}$.}

\subsection{The Fuchsian ODE for $\chi^{(3)}$.}

With an original method\footnote[8]{It is worth noting
that in the details of this method
we encountered many times the occurrence of 
generalizations of hypergeometric functions 
to {\em several variables}, extremely similar to the ones 
occurring in Feynman diagrams~\cite{Cabral}  
(Appel, Lauricella,  Kamp\'e de Feriet, ...). We will discuss
 these quite interesting questions elsewhere. 
} \cite{ze-bo-ha-ma-05} that allows us to
write $\, \tilde{\chi}^{(3)}$ as fully integrated sums, we generated
a long series of 490 terms.
Using a dedicated program we searched for the 
finite order linear differential equation
with polynomial coefficients in the variable 
$\,w\, = \,\, s/2/(1+s^2)$, by steadily increasing the order. 
We finally succeeded \cite{ze-bo-ha-ma-04} in finding
the following linear differential equation of order 
{\em seven} for $\, \tilde{\chi}^{(3)}$ with only 359 terms of our
long series:
\begin{eqnarray}
\label{fuchs}
&& \sum_{n=0}^{7}\, a_n \cdot 
{{d^n } \over {dw^n}} F(w) \, \, = \, \, \, \, \, 0
\end{eqnarray}
with :
\begin{eqnarray}
\label{defQ}
&&a_7 \, = \, \, \nonumber \\
&& w^7 \cdot  \left(1- w \right) \,\left( 1 + 2\,w \right) \,
  {\left(1- 4\,w \right) }^5\,{\left( 1 + 4\,w \right) }^3\,
  (1 + 3\,w + 4\,w^2) \cdot  P_7(w),  \nonumber \\
&&a_6 \, = \, \, w^6 \cdot {\left(1- 4\,w \right) }^4
\,{\left( 1 + 4\,w \right) }^2 \cdot P_6(w),  \\
&&a_5 \, = \, \,
w^5\cdot {\left(1- 4\,w \right) }^3\,
\left( 1 + 4\,w \right) \cdot  P_5(w), \nonumber \\
&&a_4 \, = \, \, w^4 \cdot \left(1- 4\,w \right)^2 \cdot P_4(w), \qquad \quad \quad
a_3 \, = \, \, w^3 \cdot \left(1-4\,w \right) \cdot P_3(w)\nonumber \\
&&a_2 \, = \, \, w^2\cdot P_2(w) , \qquad \quad
a_1 \, = \, \, w\cdot P_1(w) ,\qquad \quad  a_0 \, = \, \, P_0(w)\nonumber
\end{eqnarray}
where~\cite{ze-bo-ha-ma-04} : 
\begin{eqnarray}
\label{defP7}
&&P_7(w) \,\, =\,\, 
    1568 + 15638\,w - 565286\,w^2 - 276893\,w^3 \, + \, \cdots   \nonumber \\
&& \quad \qquad + 158329674399744\,w^{27}    +39582418599936\,w^{28} 
\end{eqnarray}

The polynomials $\, P_7(w), P_6(w)$ $\cdots$, $P_0(w)$
 are polynomials of degree respectively
28, 34, 36, 38, 39, 40, 40 and 36 in $\, w$, and are given 
in~\cite{ze-bo-ha-ma-04}.
The singular points of this differential equation
 correspond to the roots 
of the polynomial corresponding to the highest
 order derivative  $\,  F^{(7)}(w)$.

The linear differential equation (\ref{fuchs}) is an equation of the {\em Fuchsian
type since there are no singular points, finite or infinite, other
 than regular singular points}. With this property, using the 
Frobenius method~\cite{ince-56,Forsyth}, it is straightforward 
to obtain the critical exponents, from the indicial equations,
 in $\, w$, for each regular singular point. These
 are given in Table 1.

\begin{table}
\unitlength.5cm
\begin{tabular}{|l|l|l|}\hline
&& \\
$w=0 	$&$			s=0$&$ 				\rho=9, 3, 2, 2, 1, 1, 1$ \\
&& \\
$w=-1/4$&$			s=-1 $&$				\rho=3, 2, 1, 0, 0, 0, -1/2$ \\
&& \\
$w=1/4$&$				s=1$&$				\rho=1, 0, 0, 0, -1, -1, -3/2$ \\
&& \\
$w=-1/2$&$			1+s+s^2=0$&$			\rho=5, 4, 3, 3, 2, 1, 0$ \\
&& \\
$w=1$&$				2-s+2s^2=0	$&$		\rho=5, 4, 3, 3, 2, 1, 0$ \\ 
&& \\
$1+3\, w+4\,w^2 =0$&$		(2s^2+s+1)(s^2+s+2)=0$&$	\rho=5, 4, 3, 2, 1, 1, 0$ \\ 
&& \\
$1/w=0$&$				1+s^2=0	$&$		\rho=3, 2, 1, 1, 1, 0, 0$ \\
&& \\
$w=w_P$, 28 roots&	$s=s_P$, 56 roots &$		\rho=7, 5, 4, 3, 2, 1, 0$ \\
&& \\
  \hline 
\end{tabular}
{\bf Table 1:} Critical exponents for each regular singular point.
$w_P$ is any of the 28 roots of $P_7(w)$. We have also shown the
corresponding roots in the $s$ variable\footnote[3]{In the variable $\, 
s$ the local exponents for $w=\pm 1/4$
are twice those given.}.
\end{table}

 We have actually found
two  remarkable {\em rational and algebraic} 
solutions of (\ref{fuchs}), namely :
\begin{eqnarray}
\label{remarkable}
S_1 \, = \, \, {{w} \over {1\, -4\, w}}, \qquad \qquad 
S_2 \, = \, \,
{\frac {{w}^{2}}{ \left( 1-4\,w \right) \sqrt {1-16\,{w}^{2}}}}
\end{eqnarray}
We also found a solution  behaving like $\, w^3$,
that we denote $\, S_3$ :
\begin{eqnarray}
&&S_3 \, = \, \, 
{w}^{3}+3\,{w}^{4}+22\,{w}^{5}+74\,{w}^{6}+417\,{w}^{7}+1465\,{w}^{8}
\nonumber \\
&& \qquad \qquad +7479\,{w}^{9}  +26839\,{w}^{10}\, + \cdots 
\end{eqnarray}
and $ \, S_9$ associated with the physical solution  $\, \tilde{\chi}^{(3)}$:
\begin{eqnarray}
&& S_9 \, = \, \, {w}^{9}+36\,{w}^{11}
+4\,{w}^{12}+884\,{w}^{13}+196\,{w}^{14} \nonumber \\
&& \qquad  \qquad \qquad +18532\,{w}^{15}+6084\,{w}^{16}\,+ \cdots  \nonumber 
\end{eqnarray}
 and three solutions  $\, S_1^{(2)}$, $\, S_2^{(2)}$ and $\,S_1^{(3)}$ 
with {\em logarithmic terms}, behaving, for small $\, w$, as follows :
\begin{eqnarray}
\label{others}
&&S_1^{(2)}\, = \, \, \, \ln(w) \cdot
(S_1-4\, S_2\, +16\, S_3\, -216\, S_9)\, -32\, w^4 \cdot c_2, \nonumber \\
&&S_2^{(2)}\, = \, \, \,  
\ln(w) \cdot 
(S_2\, -2\, S_3\, +24\, S_9)\, + \, 8\, w^4 \cdot d_2, \nonumber \\
&&S_1^{(3)}\, = \, \, 3 \, \ln(w)^2 \cdot
 (S_1\, +5\, S_2\, -2\, S_3)\, \,  \nonumber \\
&&\qquad \qquad \quad - 6 \,  \, \ln(w) 
\cdot (2 \, S_3 \, -S_1^{(2)} \,-9\, S_2^{(2)} )
\,\,  -19\, w^4 \cdot e_3 \nonumber
\end{eqnarray}
where $\,c_2, \, d_2 , \, e_3$ denote functions analytical at $\, w=0$:
$\, c_2 \, \, = \, \, 1 \, +167/96 \,w\, + 2273/96 \,{w}^{2}
 +\cdots $, $ \, \, d_2 \, = \, \,
1+\, 5/2 \,w+ \,103/4 \,{w}^{2}+ \, 315/4 \,{w}^{3} +\cdots$,
and $ e_3 \, = \, \,
1+ \, 7693/456\,w
+\, 575593/11400 \,{w}^{2}
+ \, 2561473/5700 \,{w}^{3}
+\cdots$

{\bf Remark:} Besides the known regular ``Nickel-type'' 
singularities (\ref{sols}) mentioned above, we remark
 the occurrence of the roots of the polynomial $\, P_7$ of degree 28 in $w$,
which just correspond to {\em apparent singularities}~\cite{ze-bo-ha-ma-04} and
 the occurrence of the two {\em quadratic numbers}  $\,1 +3\, w +4\,w^2 \, = \,0$
which {\em are not of the form} (\ref{sols}). The two quadratic numbers
{\em are not} on the $s$-unit circle : $|s|=\, \sqrt{2}$ and  $|s|=\, 1/\sqrt{2}$.

These new quadratic singularities that are not of the ``Nickel type'' (\ref{sols}),
are {\em singularities of the Fuchsian differential equation
associated with} $\chi^{(3)}$, but {\em are not 
singularities of the (``physical'') solution} $\chi^{(3)}$ itself~\cite{zenine4} !  
More generally, for arbitrary  $\chi^{(n)}$'s, the ``non-Nickel'' singularities~\cite{Nickel-05}, are
 singularities of the Fuchsian ODE's,
but {\em does not seem to be singularities~\cite{zenine5} of the} $\chi^{(n)}$'s. 
This is a very strong motivation to get more informations on the
next $\chi^{(n)}$'s ($\chi^{(5)}$, $\chi^{(6)}$, ...), in particular their singularities
and the singularities of the associated~\cite{zenine5} Fuchsian ODE's. 

The best series analysis of the  
 $\chi^{(n)}$'s, which does not take into
 account the existence of the corresponding ODE, 
{\em will not be able to find such
 singularities}\footnote[9]{A simple Pad\'e analysis of $\chi^{(n)}$'s series
expansions will not see such ``non-Nickel'' singularities. A 
very careful diff-Pad\'e analysis, taking
into account the Fuchsian character of these ODE's, should be
 able to see such ``non-Nickel'' singularities,
when a straight simple diff-Pad\'e analysis will
 not.}. These singularities are {\em  beyond any 
series analysis of the}  $\chi^{(n)}$'s.

It is probably not necessary to underline the fundamental 
role of the understanding 
of singularities in lattice statistical mechanics, as far as 
physics is concerned. Along 
this line let us mention the recent comment by B. Nickel~\cite{Nickel-05}
on our two first papers~\cite{ze-bo-ha-ma-04,ze-bo-ha-ma-05} where he
 gives strong arguments that these 
new sets of singularities should actually (necessarily) correspond 
to {\em pinch singularities}~\cite{Nickel-05}
totally similar to the well-known {\em Landau singularities} encountered 
in  {\em Feynman diagrams}~\cite{Nickel-05,Kershaw,loc,loc2,loc3}.
Such ``Landau singularities'' can be seen as some frontier
between different perturbation regimes and their implications on 
the physics are too numerous to be listed here. As far as terminology is concerned 
we should not call these pinched ``non-Nickel'' singularities  ``Landau singularities''
but rather ``Landau-like'' singularities : they are singularities of 
a mathematical structure associated with the  $\chi^{(n)}$'s (namely the associated Fuchsian ODE)
and {\em not}  automatically singularities of the  $\chi^{(n)}$'s themselves.

\subsection{The Fuchsian ODE for $ \, \chi^{(4)}$.}
\label{fuckhi4}

 It is clear  that the four particle contribution $\tilde{\chi}^{(4)}\, $%
 is even\footnote[3]{This is also the case for any 
$\tilde{\chi}^{(2n)}$ (see for instance~\cite{nickel-00}).} in $w$.
We thus introduce, in the following, the variable $x=16 \,w^2$.
The same expansion method \cite{ze-bo-ha-ma-05} is applied with some
symmetries and tricks specific to $\tilde{\chi}^{(4)}\, $ to generate
216 terms in the variable $x$.
We succeeded in obtaining the differential equation for
$\,\tilde{\chi}^{(4)}$ that is given in $x=16 \,w^2$ by
\begin{eqnarray}
\label{fuchs4}
\sum_{n=1}^{10}\,a_{n}(x)\cdot {\frac{{d^{n}}}{{dx^{n}}}}F(x)\,\,=\,\,\,\,\,0
\end{eqnarray}
with  
\begin{eqnarray} 
\label{defQ4}
&&a_{10}= \,  -512\,{x}^{6} \left(x-4 \right) 
 \left(1-x \right)^{6} P_{10}(x),  \nonumber \\
&&a_{9}=\,  256\, \left(1-x \right)^{5}{x}^{5} P_{9}(x),
\qquad a_{8}= -384\, \left(1-x \right)^{4}{x}^{4} P_{8}(x),  \nonumber \\
&&a_{7}=\,  192\, \left(1-x \right)^{3}{x}^{3} P_{7}(x),\qquad 
a_{6}= \,  -96\, \left(1-x \right)^{2}{x}^{2} P_{6}(x),  \nonumber \\
&&a_{5}= \, 144\, \left(1-x \right)\, x \,P_{5}(x),\qquad a_{4}= -72 \, P_{4}(x),
\quad \, \,  a_{3}= \, -108 \, P_{3}(x), \nonumber \\
&& a_2=\,  -54 \, P_2(x), \quad  \qquad a_1= \, -27 \, P_1(x)
\end{eqnarray}
where $P_{10}(x),P_{9}(x)$ $\cdots $, $P_{1}(x)$ are polynomials of degree
respectively 17, 19, 20, 21, 22, 23, 24, 23, 22 and 21, 
and are given in~\cite{ze-bo-ha-ma-05b}.


\centerline{
\begin{tabular}{|l|l|l|l|}
\hline
&  &  &    \\ 
$x$-singularity & $s$-singularity & Critical exponents in $x$ &  $P$
\\ 
&  &  &    \\ 
\hline
&  &  &    \\
$0$ & $0,\infty$        & $8,3,3,2,2,1,1,0,0,-1/2$          &   $1$ \\ 
&  &  &    \\ 
$1$ & $ \pm 1$              & $3,2,1,1,0,0,0,0,-1,-3/2$       &   $3$ \\ 
&  &  &    \\ 
$4$ & $ \pm {{1}\over{2}} \pm i {{\sqrt{3}}\over{2}}$      & $8,7,13/2,6,5,4,3,2,1,0$     &   $0$ \\ 
&  &  &    \\ 
$\infty $ & $\pm i$     & $5/2, 3/2, 3/2, 1/2, 1/2, 1/2, 1/2, 0, -1/2, -1/2$          &   $3$ \\ 
&  &  &    \\ 
$x_{P}$, 17 roots & $s_{P}$, 68 roots     & $10,8,7,6,5,4,3,2,1,0$      &   $0$ \\ 
&  &  &    \\ \hline
\end{tabular}
}
\textbf{Table 2:} Critical exponents for each regular singular point. 
$P$ is the maximum power of the logarithmic terms for each singularity.
$x_{P}$ is any of the 17 roots of $P_{10}(x)$.


\subsection{Seeking for the simplest Fuchsian ODE: paradox and subtleties.}
\label{seeking}
 As far as finding the ODE's satisfied by the $\chi^{(n)}$'s is concerned
one can, of course, find several ODE's for 
a given  $\chi^{(n)}$. Actually, denoting 
${\bf L}_q^{(n)}$ the differential operator of 
{\em smallest order} $\, q$ (order seven for 
 $\chi^{(3)}$, ten for $\chi^{(4)}$, ...)
associated with a given $\chi^{(n)}$ 
(${\bf L}_q^{(n)}(\tilde{\chi}^{(n)}) \, = \, \, 0$),
the other ODE's can be seen to be associated with  differential operators
of order $ \, p$ of the form ${\bf L}_p^{(n)}\, = \, $
$\,{\bf L}_p^{q} \cdot {\bf L}_q^{(n)}, \,$$ \, p\, > \, q$.
At this step one should try to fight the prejudice that $\,{\bf L}_p^{(n)}$,
being the product of $\,{\bf L}_p^{q}\, $ 
and (the smallest order) operator $\,  {\bf L}_q^{(n)}, \,$
is necessarily more complicated than these two differential operators: 
in some cases it may actually be simpler !
 Among these various ODE's and  associated differential 
operators, some are singled-out: 
the differential operator ${\bf L}_m^{(n)}$ such that
 one {\em does not have any apparent
 singularity anymore}
is clearly singled-out.
Note that the requirement to have no apparent
 singularities, or just a polynomial corresponding to apparent
 singularities of small degree (quadratic ...),
 corresponds to ODE's such that {\em  the number of coefficients}
in a series necessary to guess the ODE, {\em is actually 
 much smaller} than the number required
to guess the differential
 operator ${\bf L}_q^{(n)}$   corresponding to the ODE of smallest order. 
However, the requirement to have no apparent
 singularity is {\em  not the optimal condition to find (guess ...)
 the ODE with the minimal number of coefficients}. The optimal order $p$ 
for guessing the (Fuchsian) ODE verified
 by a given  $\chi^{(n)}$ and its associated 
differential operator ${\bf L}_p^{(n)}$, is, in
 fact, an integer such that $ \, q \le \, p \, \le  \,  m$.

For instance, for $\, \chi^{(4)}$ one can introduce the following
differential operators $\,   {\cal L}_{m}\,= \, \, {\bf L}_m^{(4)}$:
\begin{eqnarray}
&&{\cal L}_{14} \, = \,\, {\bf L}_{14}^{10} \cdot {\cal L}_{10}, \quad  \,
{\cal L}_{13} \, = \,\,   {\bf L}_{13}^{10} \cdot   {\cal L}_{10} \\
&& \qquad {\cal L}_{12} \, = \,\,{\bf L}_{12}^{10} \cdot {\cal L}_{10}, \quad  
{\cal L}_{11} \, = \,\,{\bf L}_{11}^{10} \cdot {\cal L}_{10}
\nonumber
\end{eqnarray}
where $\, {\cal L}_{10}$ is the differential operator of smallest order
which requires 242 coefficients to be guessed; 
$\, {\cal L}_{14}$ is a differential operator 
{\em with one apparent singularity}
that requires 137 coefficients to be guessed,
$\, {\cal L}_{13}$ is the differential operator 
which requires the {\em smallest number of 
coefficients to be guessed} namely 134 coefficients (and it has a 
quadratic  apparent polynomial).
$\, {\cal L}_{12}$ requires 143 coefficients and has a
 quartic apparent polynomial.
$\, {\cal L}_{11}$ requires 161 coefficients and 
it has an apparent polynomial of degree seven.
Note that  $\, {\bf L}_{12}^{10}$ {\em is not divisible}
 by  $\, {\bf L}_{11}^{10}$
(and similarly  $\, {\bf L}_{13}^{10}$ by  $\, {\bf L}_{11}^{10}$, etc ...). 

Therefore, we see that there are, at least, three singled-out 
ODE's : the one of smallest order $\, q$, the one with no apparent
 singularity of order $\, m$, and the ``{\em optimal one to be guessed}''
which is of order $\, p$ with $ \, q \, \le \, p \, \le  \,  m$.


\subsection{Singled-out solutions of the Fuchsian ODE for $\,\chi^{(4)}$
and associated factorizations
of differential operators.}
\label{solfac}


The order ten Fuchsian differential equation (\ref{fuchs4}) has three
remarkably simple {\em algebraic solutions}, namely the constant solution, 
$\, {\cal S}_0(x) =\,  constant$,
and :
\begin{eqnarray}
 {\cal S}_1(x) = 
\, {\frac{8-12x+3x^2}{8(1-x)^{3/2}}}, \qquad
{\cal S}_2(x) = \, {\frac{2-6x+x^2}{2(1-x)\sqrt{x}}}. \nonumber 
\end{eqnarray}
These solutions correspond to solutions of
 some differential operators of order one.
Denote these three order one differential operators 
respectively $L_0$, $L_1$ and $L_2$.

A truly remarkable finding\footnote[8]{Do note that we 
had a similar, though less spectacular,
property for the differential operator of $\, \tilde{\chi}^{(3)}$: solution
 $\, S_1$ in (\ref{remarkable}) is nothing but
$\, \tilde{\chi}^{(1)}$ and is thus also a solution
 of that operator.} {\em is the following solution of
 the Fuchsian differential equation} (\ref{fuchs4}) :
\begin{eqnarray}\, 
{\cal S}_3(x) \, = \,\,\,
 {{1}\over{64}} \,x^{2}\cdot
 {_{2}}F_{1}\Bigl({{5}\over{2}},{{3}\over{2}};3; x \Bigr)
\end{eqnarray}
{\em which is nothing but the two-particle contribution to
 the magnetic susceptibility}, i.e.,
$\tilde{\chi}^{(2)}$ associated with
 an operator of order two, $N_0$ such that $N_0({\cal S}_3) = \,0$.

We do not see this result as a coincidence, but rather as {\em the emergence
of some ``Russian-doll'' structures}\footnote[4]{For instance
$\, \chi^{(3)}$ could be a solution of the differential operator
associated with $\, \chi^{(15)}$, but {\em also} of the differential operator
associated with $\, \chi^{(9)}$.} for the infinite set of 
the  $\, \chi^{(n)}$'s, and therefore as a first example of a structure
bearing on the {\em whole} susceptibility   $\, \chi$.

 The second solution of the order-two operator $N_0$
 is simply given\footnote{Or in terms
of the MeijerG function~\cite{Meijer,Bateman} as 
$\, \pi/12 \cdot {\rm MeijerG} \left( [[],[1/2,3/2]], [[2,0],[]],x \right)$ 
(see \cite{ze-bo-ha-ma-05b}). We thank P. Abbott for pointing out a missprint
in the corresponding formula given in a previous version of \cite{ze-bo-ha-ma-05b}.} by:
\begin{eqnarray}
\label{MeijerG}
\tilde{{\cal S}}_3 (x) \, = \,\,\,
 {{\pi}\over{16}} \,x^{2}\cdot
 {_{2}}F_{1}\Bigl({{3}\over{2}},{{5}\over{2}};2; 1-x \Bigr)
\end{eqnarray}
which can also be written around $\, x =\, 0$ as (${\cal S}_3 (x)$ analytic):
\begin{eqnarray}
\tilde{{\cal S}}_3 (x) =\, 
\,  \, {\cal S}_3 (x) \cdot \ln \left( x \right) + \,
 {\frac{1}{12 \, \pi}} \cdot 
\sum _{k=0}^{\infty } {x}^{k}
{\frac{d}{dk}}\, \Bigl(  
{\frac { \Gamma \left( k-1/2 \right) \Gamma \left( k+1/2 \right) } 
{ \Gamma \left( k-1 \right) \Gamma \left( k+1 \right) } }
\Bigr) \nonumber 
\end{eqnarray}
It is obvious on the simple form (\ref{MeijerG}) that this second solution
{\em is not singular}
 at the ferromagnetic critical point
$\, x=1$, but {\em is actually singular} at $\, x=0$.

We found another solution that reads 
\begin{eqnarray}
{\cal S}_4(x) &=& \, {\frac{4\,(x-2)\sqrt{4-x}}{x-1}} +
16 \ln{  {\frac{x}{(2+\sqrt{4-x})^2}}   }   \\
&&+  16 \, {\cal S}_1(x)  \cdot  \ln{g(x)} \, -  
   16\, \sqrt{x} \, {\cal S}_2(x) \, \cdot 
 {_{2}}F_{1}\Bigl({{1}\over{2}},{{1}\over{2}};{{3}\over{2}}; {{x}\over{4}} \Bigr) \nonumber 
\end{eqnarray}
with :
\begin{eqnarray}
g(x) = \,\, {{1}\over{x}} \, \Bigl( (8-9x+2x^2) \,
 +2 \,(2-x) \cdot \sqrt{(1-x)(4-x)} \Bigr)
\end{eqnarray}
The differential equation corresponding to ${\cal S}_4(x)$ has also
${\cal S}_0$, ${\cal S}_1(x)$ and ${\cal S}_2(x)$ as solutions.

Denoting by ${\cal L}_{10}$ the differential operator corresponding to
(\ref{fuchs4}), one may search for simple solution of its adjoint
 ${\cal L}_{10}^{*}$.
We found a solution corresponding to an operator 
of order two (denoted $N^{*}_8$)
which is a combination of elliptic integrals
with polynomials~\cite{ze-bo-ha-ma-05b} of degree 21
\begin{eqnarray}
{\cal S}_1^{*}(x) =\,  {\frac{x\, (1-x)^6(4-x)}{3840000 \cdot P_{10}(x)}}
\cdot \Bigl( q_1(x) \,K(x)+q_2(x) \,E(x) \Bigr) \nonumber 
\end{eqnarray}
where :
\begin{eqnarray}
K(x)={_{2}}F_{1}\Bigl({{1}\over{2}},{{1}\over{2}};1;x\Bigr), \qquad
E(x)={_{2}}F_{1}\Bigl(-{{1}\over{2}},{{1}\over{2}};1;x\Bigr). \nonumber 
\end{eqnarray}

All these solutions of (\ref{fuchs4}) can be used to build the complete
factorization of the differential operator ${\cal L}_{10}$ that we write as
\begin{eqnarray}
\label{factoriz}
{\cal L}_{10}  \,=\, \,  N_8 \cdot M_{2}
 \cdot G(L) \, =\,  M_1 \cdot N_{9} \cdot G(L)  
\, =\, \,  M_1 \cdot L_{24} \cdot G(N)    \nonumber
\end{eqnarray}
where $G(L)$ is a shorthand notation of a differential operator of order four,
factorizable in four order one operators\footnote[8]{
The order four operator  $G(L)$ has six different 
factorizations involving thirteen
differential operators of order one.} and where $G(N)$
 is an operator of order five,
factorizable in one operator of order two and three operators of order one.
$G(N)$ has 24 different factorizations involving eight differential operators
of order two, and 24 order one operators with twelve appearing in $G(L)$.
All these 36 factorizations (\ref{factoriz}) are
 given in \cite{ze-bo-ha-ma-05b}.

From the 24 factorizations of the order five differential operator $G(N)$,
six are divisible by the differential operator $N_0$. One such factorization
reads $G(N) = L_{13} \cdot L_{17} \cdot L_{11} \cdot N_0$ implying
the {\em direct sum} decomposition :
\begin{eqnarray}
{\cal L}_{10} \,\, = \,\,\, {\cal L}_{8}  \oplus N_0,
 \qquad {\cal L}_{8} \,\,\, = \,\,\,  M_2 \cdot G(L)
\end{eqnarray}


\subsection{More factorizations for the differential 
operator of $\, \chi^{(3)}$.}
\label{morefact}

Let us come back to $\, \chi^{(3)}$. 
One can change\footnote{In some of our calculations the linear combination 
$ \, \,  S_3\, -\, 12 \, S_9 $ which amounts
 to changing the $\, w^9$ coefficient in $\, S_3$ 
from $7479$ to $7467$ has also been used.} $\, S_3$ into $\, S_3^{(new)} \, $
$= \, \,  S_3\, -\, 16 \, S_9 $.
 This amounts to changing the $\, w^9$ coefficient in $\, S_3$ 
from $7479$ to $7463$ (and the following coefficients consequently).
Actually B. Nickel found\footnote[2]{Private 
communication.} that the new solution
$\, S_3^{(new)}$ is actually {\em a solution of a fourth 
order differential equation}. 

Focusing on this differential operator
one can\footnote[5]{Using for instance the 
dsolve and DFactor command of DEtools~\cite{DEtools}. See 
also~\cite{Hoeij,Help,Singer}.}
find the general solution of this order four
 differential operator, which we call $\, N_4$, which reads:
\begin{eqnarray}
\label{genersolu}
Sol \, = \, \,\,  \alpha_1 \cdot S_1 \, + \, \alpha_2 \cdot S_2\,
+ \, \, S_2  \cdot
\int {{Sol_2 } \over { S_2}} \cdot dw 
\end{eqnarray}
where $\,Sol_2$ denotes the general solution of an order two differential
operator~\cite{zenine4} $\, Z_2$:
\begin{eqnarray}
\label{defL2}
&& Z_2 \, = \, \, \,  \,{{d} \over {dw^2 }} \, - \, {{A } \over {P } } \cdot
 {{d} \over {dw }}\, + \, {{B } \over {P } }, 
 \nonumber \\
&& P=\, w \cdot \left( 1-4\,w \right)^{2} 
\left( 1+4\,w \right)  \left( 1+3\,w+4
\,{w}^{2} \right)  \left(1-w \right)  
\left( 1+2\,w \right)\cdot P_1 \nonumber \\
&& P_1=\,96\,{w}^{4}+104\,{w}^{3}-18\,{w}^{2}-3\,w+1  \nonumber \\
&& A= \,\left(1-4\,w \right)  \Bigl( 49152\,{w}^{10}+72704\,{w}^{9}-48384
\,{w}^{8}-94272\,{w}^{7} \nonumber \\
&& \quad -40368\,{w}^{6}-4488\,{w}^{5}+1080\,{w}^{4}-
108\,{w}^{3}-111\,{w}^{2}-6\,w+1 \Bigr) \nonumber \\
&& B=\,98304\,{w}^{10}+98304\,{w}^{9}-239616\,{w}^{8}-307200\,{w}^{7}-71552
\,{w}^{6}  \nonumber \\
&& \quad \quad -4224\,{w}^{5}-3192\,{w}^{4}
-1520\,{w}^{3}-276\,{w}^{2}+48\,w+4.  \nonumber
\end{eqnarray}

Let us make a first remark: the degree four polynomial $\, P_1$ 
corresponds to {\em apparent singularities} of the order two differential
operator $\, Z_2$.
This differential operator appears when solving the order four differential
operator $N_4$ which has seven apparent singularities roots of a polynomial
of degree seven.
 One thus sees that the
degree 28 polynomial (denoted $\, P_7$ in~\cite{ze-bo-ha-ma-04}, 
see equation (9) in~\cite{ze-bo-ha-ma-04}),  corresponding to 
28 apparent singularities, has been replaced, as far as $\, N_4$ is
concerned, by a degree seven polynomial or, as far as $\, Z_2$ is
concerned, by the degree four polynomial $P_1$.
The apparent singularities
of these various differential operators are clearly highly volatile.


The linear combination (\ref{genersolu}) of four simple 
solutions is in complete agreement with the (Dfactor/DEtools)
factorization of this order four differential operator $\, N_4$:
\begin{eqnarray}
\label{L7facto}
&&N_4 \, = \, \, \, \, \, \, N_3 \oplus   L_1,\qquad L_1(S_1) \, = \, 0 \\
&& N_3 \, \, = \, \, \, Z_2 \cdot N_1,  
\qquad \quad N_1(S_2) \, = \, 0. \nonumber 
\end{eqnarray}

From the existence of this
order four differential operator one immediately deduces an order three
differential operator $\, L_3$, such that
the order seven differential operator
$\,L_7$ factorizes as follows:
\begin{eqnarray}
\label{L7facto2}
L_7 \, = \, \, L_6 \oplus  \, L_1, \quad \hbox{with:} \quad
 L_6 \,= \, L_3 \cdot N_3 \, = \, \, L_3 \cdot  Z_2 \cdot N_1  
\end{eqnarray}
Let us note that the four solutions (\ref{genersolu}) 
are {\em totally decoupled} from the linear combination 
$\, S_9\, + \mu \cdot S_1$ such that
 $\,L_6(S_9\, + \mu \cdot S_1)\, = \, \, 0$. 

Let us also mention that at the singular
 points $\, w=1$, $\, w=-1/2$, and at the two quadratic roots
 $w_1$, $w_2$ of $\, 1+3w+4w^2=\, 0$, the 
solution carrying a logarithmic term is
in fact a solution of $\, Z_2 \cdot N_1$.
Therefore, the three solutions of the differential operator
$\, L_3 \cdot Z_2 \cdot N_1$, emerging from
 $\, L_3$, are analytical at the non physical singular points $w=1$, $w=-1/2$,
and at the quadratic roots of $1+3w+4w^2=0$.
The solutions of the differential operator $\, L_3$, itself,
 are of ``elliptic integral type''~\cite{zenine4}.
At the singular point $w=1/4$, one also
 notes that the differential operator $\, Z_2 \cdot N_1\, $
is responsible of the $\, (1-4w)^{-1}$ behavior.
One will  then expect the "ferromagnetic 
constant" $\, I_3^{+}$ to be localized in the blocks of the
connection matrix corresponding to the solutions of
the order three differential operator 
$\, Z_2 \cdot N_1$ at the point $\, w=1/4$.

Note that finding such complete factorization for the differential
operator $\, L_7$ (or ${\cal L}_{10}$ for $\chi^{(4)}$) is of great help 
to get the {\em differential Galois group} (which identifies
 here with the {\em monodromy group}), and, more precisely,
when computing the {\em connection matrices}
necessary to write the monodromy matrices written in the
{\em same} (Frobenius series solutions) basis.


\section{Differential Galois group, singular behaviour and asymptotics}
\label{diffGalois}
\subsection{Differential Galois group.}
\label{subdiffGalois}

From this factorization (\ref{L7facto}) of $\,L_7$ one can
 deduce, immediately, the form\footnote[5]{One has to be 
careful with the definition one takes for these 
matrices, which will be associated
with the choice of a vector set of solutions, as a line vector or as a column
vector. Different choices change these matrices into their transpose. }
in some ``well-suited global basis'' of solutions, of
 the $\, 7 \times 7$ matrices
representing the differential Galois\footnote[3]{Since our
 ODE is a Fuchsian ODE the differential Galois group identifies
with the monodromy group generated by the monodromy 
matrices written in the same basis: there are no Stokes
matrices~\cite{Sto} associated with irregular singular
 points.} group~\cite{zenine4,Moscou,Put,Alexa} of $\,L_7$, namely:
\begin{eqnarray}
\label{matgal}
 \left[ \begin {array}{ccccccc} 
{\it a1}&0&0&0&0&0&0\\
\noalign{\medskip}0&{\it b1}&0&0&0&0&0\\
\noalign{\medskip}0&{\it h1}&{\it g1}&{\it g2}&0&0&0\\
\noalign{\medskip}0&{\it h2}&{\it g3}&{\it g4}&0&0&0\\
\noalign{\medskip}0&{\it H1}&{\it H2}&{\it H3}&{\it G1}&{\it G2}&{\it G3}\\
\noalign{\medskip}0&{\it H4}&{\it H5}&{\it H6}&{\it G4}&{\it G5}&{\it G6}\\
\noalign{\medskip}0&{\it H7}&{\it H8}&{\it H9}&{\it G7}&{\it G8}&{\it G9}
\end {array} \right]
\end{eqnarray}
where the  $\, 2 \times 2$ and $\, 3 \times 3$ block matrices 
$\, {\cal G} al(Z_2) $ and $\, {\cal G}al(L_3) $
\begin{eqnarray}
\label{galgal}
 {\cal G} al(Z_2) \, = \, \,  \left[ \begin {array}{cc} 
{\it g1}&{\it g2}\\
\noalign{\medskip}{\it g3}&{\it g4}
\end {array} \right], 
\qquad 
 {\cal G}al(L_3) \, = \, \,  \left[ \begin {array}{ccc} 
{\it G1}&{\it G2}&{\it G3}\\
\noalign{\medskip}{\it G4}&{\it G5}&{\it G6}\\
\noalign{\medskip}{\it G7}&{\it G8}&{\it G9}
\end {array} \right]\nonumber 
\end{eqnarray}
correspond respectively to the differential Galois groups
 of the differential operators $\, Z_2$ and $\, L_3$.
The $\, 3 \times 3$ matrix $\, {\cal G}al(N_3)$, 
corresponding to the differential Galois group of $\, N_3 \, = \, Z_2 \, N_1$,
and the $\, 3 \times 3$ block-matrix corresponding  to 
the fact that we
have a {\em semi-direct product} 
of the differential Galois group of $\, L_3$ and $\,  N_3$ 
in $\, L_6 \, = \, L_3 \cdot N_3$,
read:
\begin{eqnarray}
{\cal G}al(N_3)\, = \, \,
 \left[ \begin {array}{ccc} 
{\it b1}&0&0\\
\noalign{\medskip}{\it h1}&{\it g1}&{\it g2}\\
\noalign{\medskip}{\it h2}&{\it g3}&{\it g4}
\end {array} \right], \quad \quad  
{\cal H}(N_3; L_3)\, = \, \,  \left[ \begin {array}{ccc} 
{\it H1}&{\it H2}&{\it H3}\\
\noalign{\medskip}{\it H4}&{\it H5}&{\it H6}\\
\noalign{\medskip}{\it H7}&{\it H8}&{\it H9}
\end {array} \right]\nonumber 
\end{eqnarray}
The semi-direct product structure is also clear on the $\, 3 \times 3$ matrix
$\, {\cal G}al(N_3)$. 
The  $\, 3 \times 3$ matrix
block $\, {\cal H}(N_3; L_3)$ (resp. the entries
 $\, h_1$ and $\, h_2$ in (\ref{matgal}))
can, in principle be calculated exactly from\footnote{In 
mathematical words the solutions of  $\,L_3$
are in the Picard-Vessiot extension of $\, L_6$.
} the 
differential Galois group of $\, N_3$, and the
 differential Galois group of $\,L_3$
(resp. the differential Galois group of  $\, L_2$ and 
the differential Galois group of $\, N_1$).
The calculation of these off-diagonal blocks, associated
 with the fact that we have a semi-direct product
instead of a product (a product of differential operators instead of a direct 
sum of  differential operators), relies on the fact
 that {\em one knowns how to actually calculate 
the morphisms between differential operators} (see 
for instance P. Berman and M. Singer~\cite{BermSing}
or  P. Berman~\cite{Berm}). It is worth noting that
 such calculations are {\em totally effective}. 
We will, not however, give more details of these later 
calculations~\cite{BermSing,Berm,Weil}: they are only necessary if one wishes
to perform exhaustively the complete analysis of
 the differential Galois group of $\, L_7$:
we just perform, here, a ``warm-up'', only sketching the analysis of 
the differential Galois group of $\, L_7$, which
 reduces (up to a product by the field  of complex
 numbers $\, \mathcal{C}$, because  $\, L_7$
is the direct sum of $\, L_6$ and $\, L_1$)
to the differential Galois group of $\, L_6$,
which is, thus, just seen as a semi-direct product 
of the differential Galois group of $\, L_3$, of the
 differential Galois group of $\, L_2$
and of the differential Galois group of $\, N_1$ (namely the field 
of complex numbers $\, \mathcal{C}$).
At first sight the differential Galois group of $\, L_3$ can be either
the group $\, SL(3, \, \mathcal{C})$, or the 
group\footnote[2]{Such a  $\, PSL(2, \, \mathcal{C})$ situation
would correspond to building differential
 equations such that one of their solutions
is actually the product of two solutions of a second 
order differentiable equation. One
 can easily build such examples using the $\, diffeq* diffeq$ 
command in Gfun~\cite{Gfun}.}  
$\, PSL(2, \, \mathcal{C})$, or the semi-direct product
of $\, \mathcal{C}$ and of the permutation group
$\, \Sigma_3$ of three elements. Some calculations that will not be detailed
here\footnote[1]{A publication with Jacques-Arthur
 Weil on these questions will 
follow~\cite{Weil2}.}, show that the differential Galois group of 
$\, L_3$ is actually $\, SL(3, \, \mathcal{C})$  
because the Wronskian of $\, L_3$ is rational 
and that the differential Galois group of $\, L_2$  is  
$\, SL(2, \, \mathcal{C})$
also because its Wronskian is rational.
The differential Galois group of $\, L_7$ thus reduces (up 
to a product by $\, \mathcal{C}$)
 to semi-direct products
of $\, SL(3, \, \mathcal{C})$,  $\, SL(2, \, \mathcal{C})$
 and $\, \mathcal{C}$. 
This is a quite general description of the differential Galois group, that may 
satisfy mathematicians, but, as far as physics is concerned, we need 
to actually calculate specific elements of the  differential Galois group,
in particular the  ``non-local'' 
{\em connection matrices}, in order to
have some understanding of the analytical properties of
 the solutions~\cite{zenine4} (dominant and sub-dominant 
singular behaviours, asymptotics, ..., see section (\ref{from}) below). 
As far as explicit calculations are concerned, we have, however, learned,
 from this quite general analysis,
that a well-suited ``global'' basis for writing 
explicitely any kind of `` non-local'' connection matrices, should be :
$\,S_1, \, \,   S_2,  \, \,   S_1^{(2;new)},  \, \, 
S_3^{(new)}$ (corresponding to the solutions of 
(\ref{genersolu}), on the $\, N_4$ side) together 
with the further solutions 
$\, S_9, \, \,  S_2^{(2)},  \, \, \,  S_1^{(3)}$ 
coming from the right division of $L_7$
by $\, N_4$.

 In particular it is shown in~\cite{zenine4}, and 
this will be sketched in the next section
 that the {\em connection 
matrix} between the singularity points $0$ and  $1/4$ (matching the well-suited
local series-basis near $\, w=0$ and the well-suited
local series-basis near $\, w=1/4$) is a matrix 
of the form (\ref{matgal}) where
 the entries are expressions in terms of $\ln(2)$, $\ln(6)$,
 $\sqrt{3}$, $\pi^2$, $1/\pi^2$, ... 
and a constant  $\, I_3^{+}$ 
introduced in equation (7.12) of~\cite{wu-mc-tr-ba-76}:
\begin{eqnarray} 
\label{Wu}
&& {{1} \over {2\, \pi^2}} \cdot 
\int_{1}^{ \infty}\int_{1}^{\infty} \int_{1}^{ \infty} 
dy_1\, dy_2\, dy_3 \Bigl({{y_2^2\, -1 }
 \over {(y_1^2\, -1)\,(y_3^2\, -1)  }}\Bigr)^{1/2} \cdot 
Y^2\, = \, \, \,\nonumber \\
&&\, = \, \, \, .000814462565662504439391217128562721997861158118508 \cdots 
\nonumber \\
&& Y \, \, = \, \,\,\, {{ y_1\, -y_3} \over { 
(y_1\, +y_2)\,(y_2\, +y_3)\,(y_1\, +y_2\, +y_3)  }}. \nonumber
\end{eqnarray} 
This constant can actually be written in term of the
 Clausen function $\,Cl_2$ :
\begin{eqnarray}
\label{I3plus}
I_3^{+} \, = \, \, {{1} \over {2 \pi^2 }}\cdot \Bigl( {{\pi^2} \over {3}} \,
 +2 \, -3 \sqrt{3}\cdot  Cl_2({{\pi} \over {3}})  \Bigr)
\end{eqnarray}
where  $\,Cl_2$ denotes  the
 Clausen function :
\begin{eqnarray}
\label{Claus}
   Cl_2(\theta) \, = \, \, 
\sum^{\infty}_{n=1}  \, {{\sin(n\,\theta) } \over {n^2 }} \nonumber
\end{eqnarray}

Similarly one can consider the (Frobenius series)
 solutions of the differential operator 
associated with $\, \chi^{(4)}$ around $\, x\, =\,0$ and around the 
ferromagnetic and antiferromagnetic critical point $\, x=1$ respectively.
Again the corresponding {\em connection 
matrix} (matching the solutions around the singularity
 points $x=0$ and the ones around the singularity point $x=1$)
have entries which  are expressions in terms of
 $\pi^2$, rational numbers but also 
of constants like constant  $\, I_4^{-}$ introduced in~\cite{wu-mc-tr-ba-76}
which can actually be written in term of the Riemann zeta
function, as follows :
\begin{eqnarray}
\label{I4moins}
I_4^{-} \, = \, \,{{1} \over {16 \pi^3 }}
\cdot \Bigl( {{4\,\pi^2} \over {9}} \, -{{1} \over {6}} 
 -{{7} \over {2}}\cdot \zeta(3)
 \Bigr) 
\end{eqnarray}
The derivation of these two results (\ref{I3plus}), (\ref{I4moins}) has
 never been published\footnote[5]{We thank C. A. Tracy for 
pointing out the existence of these two
 results (\ref{I3plus}), (\ref{I4moins}) 
and reference~\cite{Tracy}.} but 
these results appeared in a conference proceedings~\cite{Tracy}.
We have actually checked that $\, I_3^{+}$
and  $\, I_4^{-}$ we got in our calculations 
of connection matrices displayed in~\cite{zenine5}
(and in the next section) as floating numbers 
with respectively 421 digits and 431 digits accuracy,
{\em are actually in agreement} with the previous two formula. 
These two results (\ref{I3plus}), (\ref{I4moins})  provide
 a clear answer to a  question that will be addressed in
the next section of how ``complicated
and transcendental'' some of our constants occurring in the entries of the
connection matrices can be. These two
 remarkable exact formulas (\ref{I3plus}), (\ref{I4moins})
are not totally surprising when one recalls the deep link between 
 zeta functions, polylogarithms and hypergeometric 
series~\cite{Tanguy,Tanguy2,Tanguy3,Tanguy4}.
Along this line, and keeping in mind that we see all our Ising susceptibility
calculations as a ``laboratory'' for other more
 general problems (Feynman diagrams, ...),
we should also recall the various papers 
of D. J. Broadhurst~\cite{Broadhurst,Broadhurst1,Broadhurst2,Broadhurst3}
where $ \, Cl_2({{\pi} \over {3}})$ and $ \,\zeta(3) $ actually
 occur in a Feynman-diagram-hypergeometric-polylogarithm-zeta
framework (see for instance equation (163) in~\cite{Broadhurst}).

\subsection{Connection matrices and monodromy matrices.}
\label{conn}

Connecting various sets of Frobenius series-solutions
well-suited to the various sets of regular singular points
 amounts to solving a linear system of 36 unknowns
(the entries of the connection matrix). We have
 obtained these entries in floating point form with a
very large number of digits (400, 800, 1000, ...).
We have, then, been able to actually ``recognize'' these entries obtained
in floating form with a large number of digits~\cite{zenine4}.
As an axample the connection matrix for the order six differential operator $\, L_6$ 
 matching the  Frobenius series-solutions
around $\, w=0$ and the ones around $\, w=1/4$, namely $\, C(0,1/4)$
 reads:
\begin{eqnarray}
\label{C014}
&& C(0,\, 1/4)\, = \, \, \\
&& \left[ \begin {array}{cccccc} 
1&0&0&0&0&0\\
\noalign{\medskip}
1&0&-{\frac {9 \sqrt {3}}{64 \pi}}&0&0&0\\
\noalign{\medskip}
0 &-{\frac {3 \pi \sqrt {3}}{32}} &0&0&0&0\\
\noalign{\medskip}
5&{\frac {1}{3}}-2 \cdot I_3^{+}&{\frac {3 \sqrt {3}}{64 \pi}}&0&0&{ \frac {1}{16 \pi^2} }\\
\noalign{\medskip}
-{\frac{5}{4}}&-{\frac {3 \pi \sqrt {3}}{32}} &{\frac {45 \sqrt {3}}{256 \pi}}&0&{\frac{1}{32}}&0\\
\noalign{\medskip}
{\frac{29}{16}}-{\frac{2 \pi^2}{3}}&{\frac {15 \pi \sqrt {3}}{64}}
&-{\frac {225 \sqrt {3}}{1024 \pi}} -{\frac {3 \pi \sqrt {3}}{64}}&{\frac {\pi^2}{64}}&0&0
\end {array} \right]
\nonumber
\end{eqnarray}
Not surprisingly\footnote[8]{One can expect the
 entries of the connection matrices to be {\em evaluations}
of (generalizations of) hypergeometric functions,
 or solutions of Fuchsian differential equations. 
} a lot of $\, \pi$'s pop out in the entries of
 these connection matrices. We will 
keep track of the $\, \pi$'s occurring in the
 entries of connection matrices through
the introduction of the variable $\, \alpha \, = \, 2\, i\, \pi$. 

From the local monodromy matrix $\, Loc(\Omega)$,
expressed in the 
 $\, w\, = \, 1/4$ well-suited local series-basis, 
and from the connection matrix (\ref{C014}),
 the monodromy matrix around 
 $\, w\, = \, 1/4$,  expressed 
in terms of the $\, (w=0)$-well suited basis reads:
\begin{eqnarray}
\label{MalphaOmega}
&&24 \, \alpha^4 \cdot M_{w=0}(1/4)(\alpha, \, \Omega)\,\, = \, \,  
\left[ \begin {array}{cc} 
{\bf A}&{\bf 0}\\
\noalign{\medskip}
{\bf B}&{\bf C}
\end {array} \right]  \\
&&\,\qquad \quad \quad  = \, \, C(0, \, 1/4) \cdot Loc(\Omega)
\cdot C(0, \, 1/4)^{-1} \nonumber
\end{eqnarray}
where
$
\left[ \begin {array}{c} 
{\bf A}\\
\noalign{\medskip}
{\bf B}
\end {array} \right] 
$ 
and
$
\left[ \begin {array}{c} 
{\bf C}
\end {array} \right]
$ read respectively:
\begin{eqnarray}
 \left[ \begin {array}{ccc} 
-24\,{\alpha}^{4}&0&0
\\
\noalign{\medskip}
-48\,{\alpha}^{4}&24\,{\alpha}^{4}&-144\,{\alpha}^{2}\Omega
\\
\noalign{\medskip}
0&0&24\,{\alpha}^{4}
\\
\noalign{\medskip}
-48\, \rho_1 
&32\,\Omega\, \rho_2 
&48\,\Omega\, \, (9\,{\alpha}^{2}+80\,\Omega) 
\\
\noalign{\medskip}
12\,{\alpha}^{2} \rho_3 
&4 \, (75-4\,{\alpha}^{2})\,{\alpha}^{2}\Omega 
&-300\,{\alpha}^{2}\Omega
\\
\noalign{\medskip}
- \, (87+8\,{\alpha}^{2}){\alpha}^{4} 
&0&3 \, (4\,{\alpha}^{2} -75)\, {\alpha}^{2}\Omega
\end {array}
 \right],
\nonumber
\end{eqnarray}
with $\, \rho_1\, = \,  5\,{\alpha}^{4}
+8\,{\Omega}^{2}+8\,{\Omega}^{2}{\alpha}^{2} $, 
$\, \rho_2\, = \, \,4\,\Omega\,{\alpha}^{2} -75\,\Omega-15\,{\alpha}^{2}  $ and
 $\, \rho_3\, = \, 5\,{\alpha}^{2}+4\,\Omega+4\,\Omega\,{\alpha}^{2} $, 
and:
\begin{eqnarray}
C \, = \, \, \,  \left[ \begin {array}{ccc} 
24\,{\alpha}^{4}&-384\,{\alpha}^{2}\Omega&1536\,{\Omega}^{2}
\\
\noalign{\medskip}
0&24\,{\alpha}^{4}
&-192\,{\alpha}^{2}\Omega
\\
\noalign{\medskip}
0&0&24\,{\alpha}^{4}
\end {array}
 \right]
\nonumber
\end{eqnarray}
Note that the transcendental constant $\, I_3^{+}$ has disappeared in the 
final exact expression of (\ref{MalphaOmega}) {\em which actually depends
only on} $\, \alpha$ and $\, \Omega$.
This $\, (\alpha, \,\Omega)$ way of writing the 
monodromy matrix  (\ref{MalphaOmega})
enables to get straightforwardly the $\, N-$th power of (\ref{MalphaOmega}):
\begin{eqnarray}
 M_{w=0}(1/4)(\alpha, \, \Omega)^N\,\, = \, \,\, 
  M_{w=0}(1/4)(\alpha, \, N\cdot \Omega)
\nonumber 
\end{eqnarray}

Our ``global'' (800 digits, 1500 terms in the series, ...) calculations yield
quite involved exact connection matrices. With such large, and
involved, computer calculations there is always 
a risk of a subtle mistake or misprint.
At this step, and in order to be ``even more confident''
in our results, note that the monodromy matrices must satisfy one
matrix relation which will be an {\em extremely severe non-trivial check}
on the validity of these
eight matrices $\, M_i$, or more precisely their $\, (\alpha, \, \Omega)$
extensions. Actually it is known (see for instance
 Proposition 2.1.5 in~\cite{Alexa}), that 
the {\em monodromy group}\footnote[5]{Which identifies in our Fuchsian case
to the differential Galois group; we have regular singular points, and no
irregular points with their associated Stokes matrices~\cite{Sto}.}
of a linear differential equation (with $r$ regular singular points)
 is generated by a set of matrices
$\, \gamma_1, \,\gamma_2, \, \cdots, \, \gamma_r$ that satisfy 
$\, \gamma_1 \cdot \gamma_2  \cdots \gamma_r\, = \, {\bf Id}$,
where $\, {\bf Id}$ denotes  the identity matrix.
The constraint that ``some'' product of all these matrices
should be equal to the identity matrix, looks quite simple, but is, in fact,
``undermined'' by subtleties of complex analysis
on how connection matrices between non neighboring singular points
should be computed.
The most general element of the monodromy group reads:
\begin{eqnarray}
\label{g}
 M_{P(1)}^{n_1} \cdot  M_{P(2)}^{n_2} \cdot  M_{P(3)}^{n_3} 
\cdot  M_{P(4)}^{n_4} \cdot  M_{P(5)}^{n_5}
  \cdot  M_{P(6)}^{n_6}  \cdot  M_{P(7)}^{n_7} \cdot  M_{P(8)}^{n_8} \,
\end{eqnarray}
where $\, P\, $ denotes are arbitrary permutation of the 
eight first integers.
In other words, one of the products (\ref{g}) must be equal to
the identity matrix for some set of $\, n_i$'s and for some permutation $\, P$.
Let us introduce the following  choice  of ordering of the eight
singularities, namely $\,\infty , \, \, 1, \, \, 1/4, \, \, w_1,$
$ \, \, -1/2, \, \, -1/4, \, \, 0, \, \,w_2$ ($w_1\, = \, (3 \,+\, i\cdot \sqrt{7})/8$ 
and $\, w_2\, = \, w_1^{*}$
 are the two quadratic number 
roots of $\, 1+3\, w\, + 4\, w^2\, = \, 0$), the first monodromy matrix
$\, M_1$ is, thus, the monodromy matrix at 
infinity $\,{\cal M}(\infty)$, the second monodromy $\, M_2$
matrix  being the monodromy matrix at $\, w=1$, $\,{\cal M}(1)$, ...
This is actually the particular choice  of ordering of the eight
singularities, such that a
 product like (\ref{g}) is equal to the identity
 matrix\footnote[5]{Of course, from 
this relation, one also has seven other relations
deduced by cyclic permutations.}:
\begin{eqnarray}
\label{relid}
&& M_1 \cdot  M_2 \cdot  M_3 
\cdot  M_4 \cdot  M_4 
  \cdot  M_6   \cdot  M_7 \cdot  M_8 \, = \, \,\, {\bf Id} \\
&&\, \quad  = \, \,\, {\cal M}(\infty) \cdot {\cal M}(1) \cdot {\cal M}(1/4) 
\cdot {\cal M}(w_1) \nonumber   \\
&&\qquad \times {\cal M}(-1/2)
\cdot {\cal M}(-1/4) \cdot {\cal M}(0) \cdot {\cal M}(w_2) \nonumber 
\end{eqnarray}
It is important to note that these relations (\ref{relid}) 
{\em are not verified}
by the $(\alpha, \, \Omega)$ extension (like (\ref{MalphaOmega})) 
of the monodromy matrices $\, M_i$. If one consider relation
(\ref{relid}) for the $(\alpha, \, \Omega)$ extensions of the $\, M_i$'s, one 
will find that (\ref{relid}) is satisfied
 {\em only when} $\, \alpha$ {\em is equal
 to} $\, \Omega$,  but (of course\footnote[8]{A matrix
 identity like (\ref{relid}) yields a set of 
polynomial (with integer coefficients) relations on 
$\, \Omega\, = \, \, 2 \, i \, \pi$. The number $\, \pi$ being transcendental
it is not the solution of a polynomial with integer coefficients. These 
polynomial relations have, thus, to be
 {\em polynomial identities valid for any} $\, \Omega$.})
 this $\, \alpha \, = \, \Omega$ matrix 
identity {\em is verified for any value of}
 $\, \Omega$, not necessarily equal
to $\, 2 \, i \, \pi$. 

A last remark is the following:
right now, we have considered all the matrices (connection and therefore
monodromy matrices expressed in a {\em unique}
 non-local basis) with respect to the ($w=0$) well-suited basis
of solutions. This is motivated by the physical solution $\tilde{\chi}^{(3)})$
which is known as a series around $w=0$.
In fact, we can switch to another $w=\tilde{w}$ well-suited
 basis of solutions. This amounts to considering the connection
$C(\tilde{w}, w_s)=C^{-1}(0, \tilde{w})\cdot C(0,w_s)$.
For instance,  we have actually
performed the same calculations for the  $(w=1/4)$ basis of series solutions.
We have calculated all the connection matrices from  the  $(w=1/4)$ basis
to the other singular point basis series solutions, and deduced the exact
expressions of the monodromy matrices now expressed {\em in the same} 
 $(w=1/4)$ {\em basis of series solutions}. 
It is worth noting that we get, this time, for the
 monodromy $\, M_{w=1/4}(w_s)$ 
around the singular point $\, w_s$ and expressed in the  $(w=1/4)$
basis, a matrix {\em whose entries depend rationally on}
 $\, \alpha\, = \, 2\, i\, \pi$,
$\, \Omega\, = \, 2\, i\, \pi$, {\em but, this
 time, also (except for the monodromy matrix for $w=1$) on the
 "ferromagnetic constant"} $\, I_3^{+}$.
One verifies that the product of these 
$\,(\alpha, \, \Omega, \,  I_3^{+})$-dependant monodromy
matrices in the {\em same order} as (\ref{relid}),  
is actually equal to the identity matrix 
when $\, \alpha \, = \, \Omega$, the matrix identity being valid for any 
value of $\, \alpha \, = \, \Omega$ (equal or not to $\, 2\, i\, \pi$),
and {\em for any value of} $\, I_3^{+}$
(equal or not to its actual value).

We have similar results for the monodromy matrices 
around singular point $\, w_s$, expressed in the  $(w=\infty)$ basis, but,
now, the  monodromy matrices $\, M_{w=\infty}(w_s)$ depend on
$\, \alpha\, = \, 2\, i\, \pi$,
$\, \Omega\, = \, 2\, i\, \pi$, and, this time, on some (as yet unrecognized)
constants $\, y_{41}$ and $\, x_{42}$ (which can be 
found in the Appendix C in~\cite{zenine4}).
Again, the product of these monodromy
matrices in the same order as (\ref{relid}),  is
 actually equal to the identity matrix 
when $\, \alpha \, = \, \Omega$, the matricial identity being valid for any 
value of $\, \alpha \, = \, \Omega$ (equal or not to $\, 2\, i\, \pi$)
and for any values of $\, y_{41}$ and $\, \,x_{42}$.

\subsection{From the differential Galois group to 
the singularity behaviour of solutions and other
 asymptotics $ \, \chi$ studies.}
\label{from}
The knowledge of these connection matrices
gives, in particular, the decomposition of the  $\chi^{(n)}$'s
on some series well-suited to the analysis of the
neighborhood of the critical points and which, consequently, have a 
well-defined critical behaviour near the critical points. One can, then, 
immediately deduce the singularity behaviour of the  $\chi^{(n)}$'s.

Let us focus, for instance, on the critical behaviour of  $\chi^{(3)}$
near the ferromagnetic critical point $\, w \, = \, 1/4$.
Denoting $\, z\, = \, 1\, -4\, w$,
the singular part of the ``physical'' solution $\, S_9$, 
associated with $\chi^{(3)}$,
reads : 
\begin{eqnarray}
&& \tilde{\chi}^{(3)}(singular)  \, \, = \,\,  \,  8\cdot 
S_9(singular) \, \, \, \simeq  \,  \,\, \, \, 
{{ I_3^{+}} \over { 2\, z}} \, \, + \ln(z) \cdot Sing_1 \nonumber \\
&&\qquad \quad \quad \quad \quad  -{{ 1}\over {64 \, \pi^2}} \cdot 
\Bigl( \ln(z)^2 \, -6 \,\ln(2) \, \ln(z)\Bigr) \cdot Sing_2 \nonumber 
\end{eqnarray}
where 
\begin{eqnarray}
&& Sing_1 \, = \, \,\,  \,23/3\, +227\, z/240\, -2047\, z^2/1680\, 
-88949\, z^3/92160\, + \cdots \nonumber \\
&& Sing_2 \, =\, \,  \,\,  1\,-z/8 \, +3\, z^2/16\, +\, 29\, z^3/512\,
 + \cdots\nonumber 
\end{eqnarray}
and where  $\, I_3^{+}$ is {\em actually the 
transcendental constant} (\ref{Wu}),
namely $\, I_3^{+}\, = \, \, 0.000814462 \cdots \, $ 
These results agree with previous results of B. Nickel (see for instance the
footnote of page 3904 in~\cite{nickel-99}), but the correction 
terms are new\footnote[3]{All these results
have also been found by B. Nickel (private communication).
}, in particular 
the terms in $\, \ln(2)/(64 \, \pi^2)$.  
In the $\, \tau\, = (1/s-s)/2 $ variable introduced 
in~\cite{or-ni-gu-pe-01b,or-ni-gu-pe-01}
this reads 
\begin{eqnarray}
s^{1/2} \cdot \chi^{(3)} \, \, \, 
\simeq \quad 
\, 2^{1/2} \cdot \tau^{1/4}\cdot \Bigl({{I_3^{+}} \over {  \tau^2}}\, 
- {{ \ln(\tau)^2} \over {16\, \pi^2}} +\,  \cdots  \Bigr) \nonumber 
\end{eqnarray}
where the dots denote less divergent terms. 
Near the  antiferromagnetic critical point $\, w \, = \, -1/4$
we get :
\begin{eqnarray}
\tilde{\chi}^{(3)}(singular)  \,\, \simeq  \,\,\, \, \quad
{{-1} \over {32 \pi^2}} \cdot \ln^2(1+4\, w) \, + \, \cdots 
\end{eqnarray}

From these results, and using
some identities like\footnote[8]{
The asymptotic expansion of the coefficients of  Taylor series can be obtained
 using various packages availables 
at http://algol.inria.fr/libraries/software.html.
For instance one can use the command ``equivalent''
 in gfun~\cite{Gfun} (see~\cite{BSalvy}
for more details. For more details on calculations 
of asymptotics see~\cite{Flajolet,Flajolet2,wim-zei-85}.} 
\begin{eqnarray}
&&\ln^2(1-x) \, = \, \,
\sum_{n=2}^{\infty} \,  {{2} \over {n}} \cdot \Bigl(\psi(n)
 \, + \,\gamma  \Bigr) \cdot  x^n
\nonumber
\end{eqnarray}
 where $\, \gamma $ denotes the Euler Gamma function,
and $\, \psi(n)$ is the logarithmic derivative of the $\, \Gamma$
 function, and recalling,  that, when $\, n \, \rightarrow \, \infty$,
one has the asymptotic behaviour
$ \psi(n) \rightarrow \ln(n) $,
one can find the following asymptotic behavior for the
 expansion coefficients $c(n)$
of  $\tilde{\chi}^{(3)}/8/w^9 \, $ near $w=0$ :
\begin{eqnarray}
&&{{ 2} \over {4^8 \cdot 4^n }} \cdot
c(n) \,\, \,  \simeq \, \,\quad  \, {{ I_3^{+}} \over { 2}}\, -{{1} \over
 { 32 \pi^2}} \cdot {{\ln(n) } \over {n}}  + \,{{b_1} \over {n}} \nonumber \\
&&\,\qquad \qquad \qquad \quad \quad \quad
+ \,(-1)^n \cdot \Bigl( {{b_2} \over {n}} \,-{{1} 
\over { 16 \pi^2}} \cdot  {{\ln(n) }
\over {n}}\,  \Bigr)\, +\,
\cdots\nonumber
\end{eqnarray}
with  $\, b_1\,  \simeq \, 0.0037256 \,  $ and
$\, b_2\,  \simeq \, 0.0211719 \,  $ .
This parity behavior explains why we had so much
 difficulty to actually find
the ``true'' asymptotic behavior of these coefficients after obtaining around
$\, 500$ coefficients (see equations (3.6) in~\cite{ze-bo-ha-ma-05}:
one has $13.5 \times 4^N$ for $\, N$ even and  $11 \times 4^N$
for $\, N$ odd,
instead of the ``true''
 $ \,\, \,  (4^7 \cdot I_3^{+}) \times 4^N \,  \simeq \, 13.4 \times 4^N$
 asymptotic behavior).
\subsection{Generalization to $\chi^{(4)}$}
\label{khi4}
All the previous calculations can be performed, {\em mutatis mutandis},
for  $\chi^{(4)}$. For instance from the singular behaviour near $x=1$ :
\begin{eqnarray}
\tilde{\chi}^{(4)}(singular) 
\, \, \, \simeq \, \,  \, \, \, \,
{{ I_ 4^{-} } \over {1-x}}
 \, + {{ 1} \over {384\, \pi^3}} \cdot
\ln^3(1-x)\, + \,\,
\cdots
\end{eqnarray}
where $\,  I_ 4^{-}$ is given by (\ref{I4moins}) ($\,  I_ 4^{-} \,
 = \, \, 0.00002544851106586 \cdots \,\, $),
one obtains the following asymptotic form for the series 
coefficients\footnote[8]{
More generally terms like $\, \ln^i(1-x) \, $ give
coefficients growing asymptotically like $\, (-1)^i (\ln(n))^{(i-1)} /n \, $.}
$ \, c(n)$ of $\tilde{\chi}^{(4)}$ near $x=0$ :
\begin{eqnarray}
c(n) \,\, \,\, \simeq \,\, \,\,\, I_ 4^{-} \, -{{ 1} \over {128\, \pi^3}}
 \cdot {{\ln^2(n)} \over {n}}
 \, + a_2\cdot {{\ln(n)} \over {n}}\,  + \, {{a_3} \over {n}}\, + \,
\cdots
\end{eqnarray}
where $\, a_2 \, \simeq \, 0.0012515\, $ and 
$\,a_3 \, \simeq \,-0.0021928$. 

\subsection{Non generic character of $\chi^{(4)}$}
\label{khi4nongener}
  As we remarked in~\cite{ze-bo-ha-ma-05}
the critical behavior of $\chi^{(3)}$ near
the ferromagnetic point $\, w \, = \, \, 1/4$, {\em does not corresponds to
the ``dominant'' exponent}
 one obtains in the set of indicial exponents
of the indicial equation at  $\, w \, = \, \, 1/4$.
The same situation occurs exactly for $\chi^{(4)}$ as will be seen below.
The singular behavior of the solutions of the differential equation
can be read easily from Table 2.
Near $x=4$, they are $\, t^{13/2}$, where $t=4-x$.
Near $x=\infty$, they are $t^{-1/2}\, \log^k(t)$, (with $k=0,1$),
$t^{1/2}\, \ln^k(t)$ (with $k=0,1,2,3$). $t^{3/2}\, \ln^k(t)$
(with $k=0,1$) and $t^{5/2}$,  where $t=1/x$.
Near $x=1$, they are $t^{-3/2}$, $t^{-1}$, $ \ln^k(t)$ (with $k=1,2,3$)
and $ t\, \ln(t)$, where $t=1-x$.
Note that these behaviors
take place for the {\em general} solution of the
 differential equation (\ref{fuchs4}).

As far as the {\em physical} solution is concerned, the dominant singular
behavior at $x=4$ (namely $t^{13/2}$), and $x=\infty$ (namely
$t^{-1/2}\,\ln(t)$),
are present in the physical solution
 $\tilde{\chi}^{(4)}$ confirming~\cite{nickel-00,or-ni-gu-pe-01b}.
At the ferromagnetic point ($x=1$ which also corresponds 
to the anti-ferromagnetic
point for $\chi^{(4)}$), with the dominant behavior $t^{-3/2}$,
the growth of the coefficients would be
$\left( 3/2 \right)_N/N! \sim \sqrt{N}$. Since this is not the case,
the coefficients of the series for large values of $N$ behaving like
$C_N \, \sim \, 0.2544 \, \cdot \,10^{-4}$, 
this $\, t^{-3/2}$ singular behavior {\em does not contribute} to the
physical solution. Only the {\em subdominant singular behavior
is actually present}.
 
As far as singularity analysis in lattice statistical mechanics
and, more specifically, as far as the 
analysis of singular behavior near a physical 
singularity, like the ferromagnetic singularity $\, x \, = \, 1$, is concerned,
we  probably have a prejudice
which consists in believing that ``physical solutions'' will correspond
to generic solutions and, thus, will
take into account the strongest possible singularity if 
a set of exponents is available (from the indicial equations).
In fact the  ``physical solution'' $\, \chi^{(4)}$ chooses to ``avoid'' this 
dominant singular behavior, and will, near 
the ferromagnetic singularity $\, x \, = \, 1$, 
correspond to ``sub-dominant''
 singular behavior. 

As far as asymptotic calculations are concerned, there are many tools
 and methods~\cite{Flajolet,Flajolet2,wim-zei-85}
 which give (after some work) asymptotic results for the ``generic'' 
solutions of an ODE (or of a recursion). 
In the ``sneaky'' situation where physics does not correspond to the 
 ``generic'' solutions associated with the strongest possible singularities,
all the asymptotic calculations become much more difficult. 
In fact such calculations really correspond to this
 difficult class of ``global''
problems of finding a ``non-local connection matrix''. 

\subsection{$\tilde{\chi}^{(1)}+\tilde{\chi}^{(3)}$
versus $\tilde{\chi}$ at scaling.}
\label{scaling}
Thus far we have sketched some mathematical aspects of
the solutions to the Fuchsian differential equations for $\tilde{\chi}^{(3)}$
and $\tilde{\chi}^{(4)}$.
However, the physics implications of the solutions we have obtained
call for some remarks near the physical critical points.
Taking, as an example, the ferromagnetic singularity for $\tilde{\chi}^{(3)}$,
the sum of the first two $n$-particle terms behave near
 $ \,\tau \, = \, \, (1/s-s)/2\, \, \simeq \, 0 \,$ as:
\begin{eqnarray}
\label{chi1chi3tau}
\tilde{\chi}^{(1)}+\tilde{\chi}^{(3)} \,& \simeq & \,\,  
{{1+I_3^{+}} \over {  \tau^2}}\,  \,
- {{ \ln^2(\tau)} \over {16\, \pi^2}}\, \,
 +  \left( \ln(2) \,-{{23}\over{24}} \right)
\cdot  {\frac{\ln(\tau)}{4\,\pi^2}}  \\
&&
+{\frac{11}{48}}+{\frac{3}{8}}I_3^{+}-{\frac{1}{4\pi^2}}
\left( \ln^2(2)-{{23}\over{12}}\ln(2)+{{14}\over{144}}  \right) 
+\cdots \nonumber 
\end{eqnarray}
The exact susceptibility, as reported in [16], yields for the
normalized susceptibility $\, \tilde{\chi} $ near
 $ \,\tau \, = \, \, (1/s-s)/2\, \, \simeq \, 0 \,$:
\begin{eqnarray}
\label{chitau}
&&\tilde{\chi} \, =\,\,\,  {{s} 
\over {(1-s^4)^{1/4}}} \cdot \chi \,\,  =\,\,\, 
 {\frac{ \left(\tau+\sqrt{1+\tau^2}\right)^{-1/2}}
{(1+\tau^2)^{1/8}}} \times  \\
&& \quad \quad \quad \Bigl(
c_1\, \tau^{-2} \,F_{+}(\tau) \, +{\frac{\tau^{-1/4}}{\sqrt{2}}}\,
\sum_{p=0}^{\infty}\sum_{q=p^2}^{\infty}\,
 b_{+}^{(p,q)} \cdot  \tau^q\, \ln^p(\tau) \Bigr)
\nonumber
\end{eqnarray}
where $c_1=1.000815260 \cdots $ is given 
with some 50 digits in \cite{or-ni-gu-pe-01b}.
$F_{+}(\tau)$ and $b_{+}^{(p,q)}$ are given in \cite{or-ni-gu-pe-01b}.
The constants $1+I^{+}_3$ and $c_1$ verify
 $\, 1+I^{+}_3+I^{+}_5\, =\, c_1$ with nine
digits, $I^{+}_5$, corresponding to $\chi^{(5)}$, is a constant given 
in \cite{wu-mc-tr-ba-76} (and with some 30 digits
in \cite{nickel-99}).
Thus, and as suggested in \cite{wu-mc-tr-ba-76}, the
 partial sums of the $\chi^{(n)}$ should
converge rapidly  to the full $\chi$.
Furthermore, adding $\chi^{(3)}$ term has resulted in a series expansion
that reproduces the first 24 terms of $\chi$ to be
 compared with only eight first terms for
$\chi^{(1)}$ series.

However, equation (\ref{chitau}) shows a $\, \tau^{-1/4}$ 
divergence as an overall factor
to the logarithmic singularities which is absent in (\ref{chi1chi3tau}).
The same situation occurs for the low temperature regime when we compare
the first two $n$-particle terms ($\tilde{\chi}^{(2)}$ 
and $\tilde{\chi}^{(4)}$)
with the full $\tilde{\chi}$ at scaling \footnote{For the leading amplitude,
$\tilde{\chi}^{(2)}$ and $\tilde{\chi}^{(4)}$ give
$1/12\pi + I_4^{-} \simeq 1.0009593\cdots /12\pi$ which is very close to
$1.0009603\cdots /12\pi$ for the full 
$\tilde{\chi}$ \cite{nickel-99}. }. This observation raises
 several profound issues: how does the 
logarithmic terms in the entire sum add up to make the
$\tau^{-1/4}$ divergence be factored out ?
If one assumes that the other $\tilde{\chi}^{(2n+1)}$ 
terms share the same singularity
structure as $\tilde{\chi}^{3}$, in particular the 
occurrence (in variable $\tau$ or $s$)
of only {\em integer} critical exponents at the ferromagnetic critical point,
the $\tau^{-1/4}$ divergence, as an overall factor, implies the
following correspondence :
\begin{eqnarray}
\sum_{n=1}^{\infty} \sum_{m=0}^{N(n)} \alpha_{n,m} 
\cdot S_{n,m} (\tau) \, \ln^m(\tau)
 \quad \rightarrow \quad 
\tau^{-1/4} \cdot
\sum_{p=0}^{\infty}\sum_{q=p^2}^{\infty}\, b_{+}^{(p,q)} 
\, \tau^q\, \ln^p(\tau)
\nonumber
\end{eqnarray}
where $S_{n,m} (\tau)$ is analytical at $\tau=0$, the 
$\alpha_{n,m}$'s are  numerical coefficients
and  $\, N(n)$ is the maximum power of logarithmic terms 
occurring in the solution around 
the ferromagnetic point of the differential equation of 
$\, \tilde{\chi}^{2n+1}$.
This correspondence requires probably a {\em very particular structure}
in the successive differential equations.
Note that the discrepancy phenomenon we discuss here may be more widespread
than that observed here. 
The challenging problem one faces here is to link {\em linear 
and non linear} descriptions of a physical problem, namely the
description in terms of an infinite number of holonomic (linear)
expressions for a physical quantity of a non linear nature.
The difficulty to link holonomic versus 
non-linear descriptions of physical problems is typically
the kind of problems one faces with the Feynman diagram approach of particle
physics.


\section{Conclusion}
\label{conclu}

Most of the results we have displayed in this paper are, in large
part, mathematical ``subtleties'' associated with the 
structure of the differential
 operators associated with the $\tilde{\chi}^{(n)}$'s 
and also surprising results
like the fact that the Fuchsian ODE's of
 $\tilde{\chi}^{(3)}$ (resp. $\tilde{\chi}^{(4)}$)
have simple {\em rational and algebraic solutions}, 
together with previously encountered
solutions  $\tilde{\chi}^{(1)}$ (resp. $\tilde{\chi}^{(2)}$).

In a miscellaneous list of these mathematical ``subtleties''
let us recall the occurrence of a large set of apparent singularities 
for the ODE of smallest order (or the fact that the 
simplest, and easiest to be guessed, ODE
is {\em not} the one of smallest order), the 
existence of ``Landau-like'' pinched singularities,
(that are not of the ``Nickel form'' (\ref{sols})), that are 
singularities of the ODE {\em but not singularities of the
``physical'' solutions} $\, \chi^{(n)}$, the fact 
that the  ``physical'' solutions
 $\tilde{\chi}^{(n)}$  are not ``generic'' solutions 
but ``mathematically singled-out''
ones (subdominant critical behavior\footnote[3]{At the ferromagnetic critical 
point $\, w=1/4$, the behavior $\, (1-4w)^{-3/2}$ {\em corresponding
to the largest critical exponent for the ODE
 is actually absent in the physical
solution} $\, \chi^{(3)}$.}, no singular behavior for the previous 
``Landau-like'' pinched singularities), ...

The fact that  $\tilde{\chi}^{(2)}$ is a solution of the differential operator
associated with $\tilde{\chi}^{(4)}$ is probably the {\em most important 
result}, since it is a strong indication of
 some ``Russian-doll-like'' mathematical
structure valid for the {\em infinite set} of differential operators
associated with all the infinite sequence of $\, \chi^{(n)}$'s.
 
It is worth noting that almost all these
 mathematical structures, or singled-out properties,
are far from being specific of the two-dimensional 
Ising model : they also occur on 
many problems of lattice statistical mechanics or, 
even\footnote[9]{The wronskian
of the corresponding differential equation is also rational, the
associated differential operator factorizes in a way 
totally similar to the one described
in sections (\ref{solfac}), (\ref{morefact}), large 
polynomial corresponding to apparent singularities
also occur, ...},  as 
A. J. Guttmann et al saw it recently, on enumerative 
combinatorics problems like, for instance, the 
generating function of the three-choice polygon~\cite{Three}.

With this specific study
of the susceptibility of the Ising model, 
 we think that, in the future, we should contemplate a ``collision'' between 
various mathematical domains and structures, namely
the theory of {\em isomonodromic deformations} (Painlev\' e
 equations and their generalizations
with unit circle natural boundaries like Chazy III 
equations, see~\cite{chazy,chazy2}, or
 Garnier systems\footnote[2]{The occurence of 
hypergeometric functions of several complex variables in the
Garnier systems is worth noting (see page 10 and page 111 paragraph 31
 in~\cite{Garnier}).}~\cite{Garnier}), the 
theory of holonomic systems
(Fuchsian equations and in particular the
 so-called {\em rigid local systems}~\cite{darmon} and 
their geometrical interpretations~\cite{Katz}),
the theory of modular forms~\cite{Halphen,Ohyama} 
(Eisenstein series~\cite{Hankel,Eisen}, Heegner
 numbers~\cite{Fond}, complex multiplication for 
elliptic curves~\cite{Fond}, modular
 equations~\cite{Ramanu3,Hanna,Ramanujan} ...),
 and various ideas related to the analysis
of Feynman diagrams (Landau singularities, generalizations of hypergeometric
functions to several complex variables, Appel functions,
 Kamp\'e de F\'eriet, Lauricella-like functions, 
polylogarithms~\cite{Broadhurst,Broadhurst1,Broadhurst2,Broadhurst3}, Riemann 
zeta functions, multiple zeta values, ...).
We also think that such ``collisions'' of concepts and structures
are {\em not} a consequence of the free-fermion 
character of the Ising model, and that
similar convergence should also be encountered on more complicated 
Yang-Baxter integrable models\footnote{The comparison 
of the Riemann zeta functions
equations obtained for the
XXX quantum spin chain~\cite{Korepin} with 
the evaluations of central binomial 
 in~\cite{Broadhurst1} 
provides a strong indication in favor of similar structures
on non-free fermion Yang-Baxter integrable models.}, the two-dimensional 
Ising model first popping out as a consequence of 
its simplicity. 
An open question is what could remain of these related structures 
for less Yang-Baxter integrable problems such as the calculations
of Feynman diagrams ? Apparently a lot seems
 to remain~\cite{Broadhurst,Broadhurst1,Broadhurst2,Broadhurst3}.
Another open problem which requires to be addressed is the (apparent ?)
scaling discrepancy, we sketched in section (\ref{scaling}),
 between the holonomic approach and more Painlev\'e oriented
approach.  

\hskip 2cm

\textbf{Acknowledgments :}
 One of us (JMM) thanks Prof. A.J. Guttmann for his interest in our
 work, his extremely valuable advice  and for 
his constant friendly support during the last decade, and more recently
for an invitation in the ARC Centre of Excellence
 for Mathematics and Statistics of
Complex Systems in Melbourne where this paper has been completed. 
 We thank Jacques-Arthur Weil for valuable
comments on differential Galois groups, and connection matrices.
We would like to thank B. Nickel for his inspired comments
on solution $\, S_3$. We thank C. A. Tracy for pointing out
the results of reference~\cite{Tracy}.
 We thank J. Dethridge for a highly optimized C++ 
program that gave an additional and very strong
 confirmation of our $\, \chi^{(4)}$
results.  We would like to thank A. J. Guttmann, I. Jensen, and W. Orrick
for a large set of useful comments on the singularity behaviour 
of physical solutions. (JMM) thanks R.J. Baxter
 for great hospitality and discussions
when visiting the ANU (Canberra) and B. M. McCoy 
for his interest in our work and so
many helpfull comments.
(S. B) and (S. H) acknowledge partial support from PNR3-Algeria.


\end{document}